\documentclass[journal]{IEEEtran}
\usepackage{color}
\usepackage{graphicx}
\usepackage{tabu,bm}
\usepackage{cite}

\ifCLASSINFOpdf
  
\else

\fi

\begin{document}

\title{Towards Intelligent Communications: Large Model Empowered Semantic Communications}

\author{Huiqiang Xie,~\IEEEmembership{Member,~IEEE,} Zhijin Qin,~\IEEEmembership{Senior Member,~IEEE,} \\Xiaoming Tao,~\IEEEmembership{Senior Member,~IEEE,} and Zhu Han,~\IEEEmembership{Fellow,~IEEE}
\thanks{Huiqiang Xie is with College of Information Science and Technology, Jinan University, Guangzhou, China (e-mail: huiqiangxie@jnu.edu.cn).  }
\thanks{Zhijin Qin and Xiaoming Tao are with the Department of Electronic Engineering, Tsinghua University, Beijing, China (e-mail: qinzhijin@tsinghua.edu.cn, taoxm@tsinghua.edu.cn).  (\textit{Zhijin Qin is the corresponding author.}) }
\thanks{Zhu Han is with the Department of Electrical and Computer Engineering, University of Houston, Houston, USA (e-mail: zhan2@uh.edu).  } 
\thanks{}}

\maketitle

\begin{abstract}
Deep learning enabled semantic communications has shown great potential to significantly improve transmission efficiency and alleviate spectrum scarcity, by effectively exchanging the semantics behind the data. Recently, the emergence of large models, boasting billions of parameters, has unveiled remarkable human-like intelligence, offering a promising avenue for advancing semantic communication by enhancing semantic understanding and contextual understanding. This article systematically investigates the large model-empowered semantic communication systems from potential applications to system design. First, we propose a new semantic communication architecture that seamlessly integrates large models into semantic communication through the introduction of a memory module. Then, the typical applications are illustrated to show the benefits of the new architecture. Besides, we discuss the key designs in implementing the new semantic communication systems from module design to system training.  Finally, the potential research directions are identified to boost the large model-empowered semantic communications.
\end{abstract}

\IEEEpeerreviewmaketitle

\section{Introduction}
In the past decade, we have witnessed the success of deep learning (DL) in fostering various industries to improve productivity and revolute the paradigm. Within the realm of communications, particularly noteworthy is the rise of DL-enabled semantic communications \cite{qin2022semantic}. It has shown the ability to improve spectral efficiency, transmission rates, energy efficiency, and overall system robustness, which has also been identified as one of the core techniques for the sixth generation (6G) and beyond. The power of DL-enabled semantic communications is rooted in the strong ability of semantic representation and semantic understanding, allowing more efficient semantic exchange for data generation and task execution. Therefore, achieving powerful semantic representations and semantic understanding is one of the key problems in improving the performance of semantic communication systems. 

The scaling laws \cite{kaplan2020scaling} empirically show that scaling up neural networks is an effective way to improve the capacity for semantic representation and semantic understanding. {By scaling up the number of parameters to billions and feeding the vast data, large models appear to show a human-like ability in intelligent tasks,} which has been proposed as a promising new technology for achieving general artificial intelligence. The remarkable milestone is the launch of ChatGPT~\cite{openai2023gpt4} which shows the surprising ability to understand contextual conditions, generate multimodal data, analyze data, etc. In sharp contrast to the existing small models, the large models can solve zero-shot or few-shot tasks through in-context learning, generate controllably multimodal data through prompt engineering, split a complex problem into multiple simple problems through the chain of thought, and simulate the specified scenarios through role play. Besides, the artificial intelligence generated content (AIGC)~\cite{cao2023comprehensive, RadfordKHRGASAM21, thoppilan2022lamda} also benefits from the large models, leading to more realistic and high-quality content generation. These applications of large models can significantly enhance the performance of semantic communication systems in terms of system training, data generation, and task execution.

\begin{figure*}[!t]
    \centering
    \includegraphics[width=120mm]{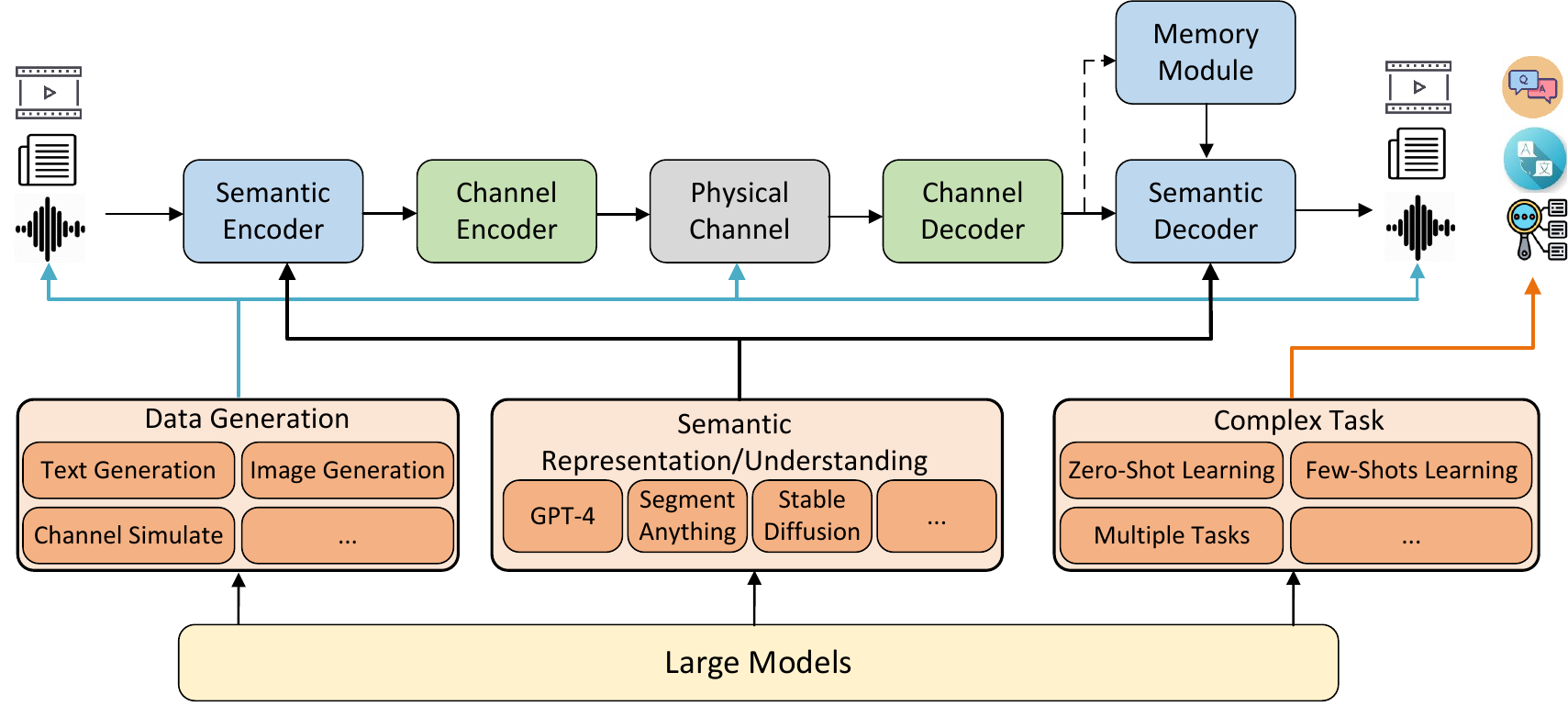}
    \caption{The proposed large model-aided semantic communication architectures.}
    \label{fig:system-model}
\end{figure*} 
Next, we elaborate on the shortcomings of existing semantic communication in achieving these applications. Review the current semantic communications for different modalities \cite{XieQLJ21, WengQ21, DaiWTSQ0022, 9837870}, given the source data, the transmitter encodes it to the transmitted data by the semantic encoder and channel encoder, and the receiver reconstructs the data or performs tasks by semantic decoder and channel decoder. In this architecture, a knowledge base is required to provide common knowledge on both sides, e.g., the training dataset, to reach a consensus. However, such a knowledge base is limited and cannot support contextual understanding in large models. Large models can employ contextual information (short-term knowledge) to understand the users' intentions, execute complex tasks, and help to generate multimodal data. Besides, large models can provide more abundant information to the system. Therefore, differing significantly from the semantic communications comprising semantic codec and channel codec only, a new large model-empowered semantic communication architecture is required.

This article introduces the new semantic communication architecture to integrate large models and conventional semantic communications and will answer these questions: \textit{Q1) What will large models bring to semantic communications? Q2) How to integrate large models into semantic communications? Q3) What are the key designs in the large model-empowered semantic communications?}  The key features and contributions of this article can be summarized as follows: 
\begin{itemize}
    \item A novel semantic communication architecture is proposed to fully utilize the power of large models by introducing the memory module, which is used to provide contextual information to the large models.
    \item The typical applications of large model-empowered semantic communication are discussed, including intention understanding, multimodal generation, complex task execution, and scenario adaptation. The large models enable semantic communications to support more types of tasks and generate diverse data.
    \item The key designs of the novel architecture are elaborated from the module design to system training, which consists of the design of the memory module, the structures of large models, the choice of joint design, the adaptive transmission to avoid outage, and the training method.
\end{itemize}

\section{New Semantic Communication Architecture}\label{sec-ii}
Fig. \ref{fig:system-model} shows the proposed semantic communication architecture. This architecture consists of not only the modules in existing semantic communications, i.e., the semantic codec and channel codec, but also the new modules, i.e., memory module and large model. {The semantics is defined as the information related to the content/tasks in the proposed architecture.}

\subsection{Contextual Information}
In our daily lives, the contextual information may refer to the dialog. However, in semantic communications, the contextual information is not limited to the dialog but includes the transmitted semantic information and the self-sensing information. The transmitted semantic information contains the multimodal semantics transmitted from the transmitter over the past time slots, e.g., text features, speech features, and visual features. This information could provide complementary information to avoid semantic ambiguity. For example, given the sentence ``I want that, instead of this,''  it is hard to understand the meaning of ``that'' in the sentence without the context ``That is apple and this is banana.''

Another kind of contextual information is the self-sensing information, e.g., user's behaviors, the environment information obtained from the cameras, and channel state information (CSI) estimated from pilots, which can provide the context for the sensing tasks, including gesture sensing and indoor location. For example, the indoor pedestrian trajectory prediction can be refined with the previous trajectory and the user's behaviors.

\begin{figure*}[!t]
    \centering
    \includegraphics[width=160mm]{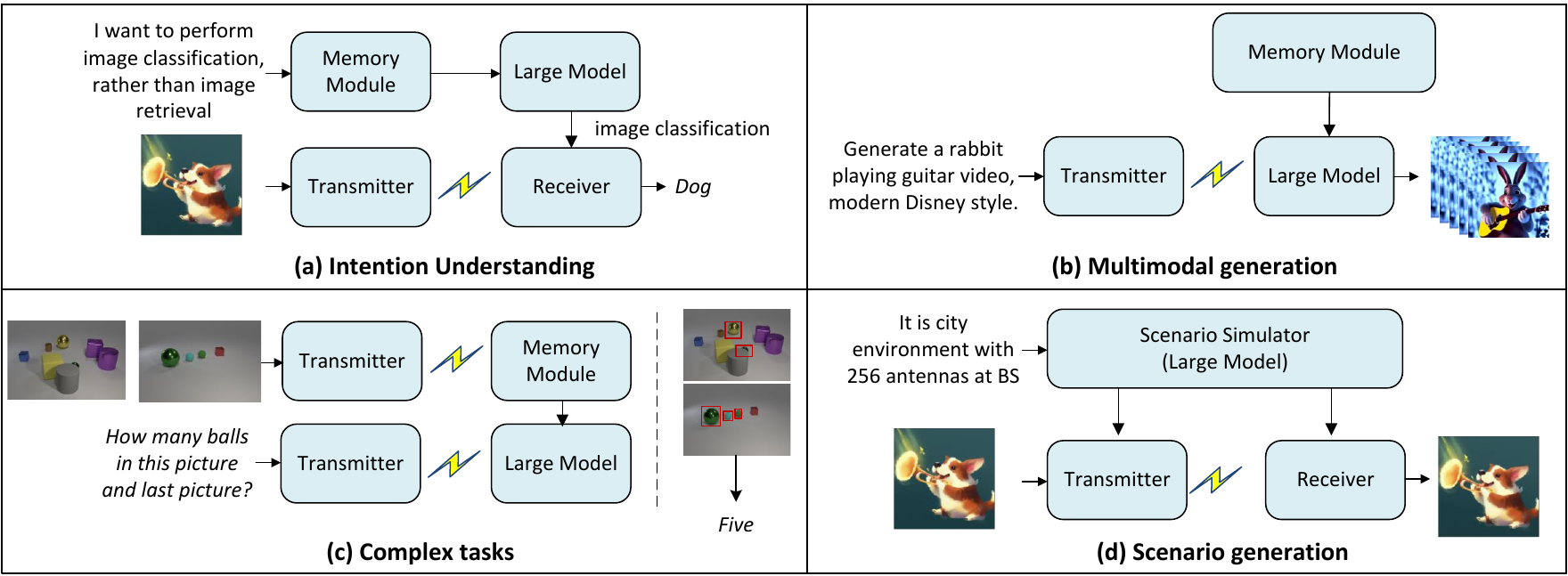}
    \caption{ Typical applications of large model-empowered semantic communications.}
    \label{fig:applications}
\end{figure*}

\subsection{Memory Module}
{The human brain has a memory area, in which humans can learn and analyze past information to help make decisions for the current situation. Besides, the large model can solve zero-shot or few-shot tasks through in-context learning. Inspired by that, we introduce the memory module in the new architecture.}  The memory module is used to store the received contextual information, which can process the stored contextual information and selectively update the stored contextual information with the received contextual information, where the out-of-date context will be replaced with fresh ones. {It is helpful for task-oriented semantic communications to support more kinds of tasks and increase the accuracy of tasks.}

Different from the large model provides general information, the memory module is the short-term knowledge. This historical information introduces the \textit{temporal domain} to semantic communications, such that the system can learn the characteristics from the few past time slots and predict the future semantics accurately.


\subsection{Large models}
Large models are the core part of powering up the capacity of semantic communications due to the diverse data generation and the strong ability of semantic representation/understanding. The large models can basically be divided into two categories according to the applications: the discriminative model and the generative model. The discriminatively large model can extract concise semantic information accurately when giving complex inputs, which can enhance the semantic understanding of semantic communications. In contrast, the generatively large model can generate data containing fine-grained details with concise semantic information. This is beneficial for semantic representation in semantic communications. By exploiting both types of large models, it is possible to enable semantic communication systems to transmit fewer semantic information but generate more complex data and serve the more difficult tasks.

\section{Potentials of Large Model-empowered Semantic Communications}\label{sec-iii}
This section introduces several typical applications of the new semantic communication architecture, including intention understanding, multimodal generation, complex task execution, and scenario generation. 

\subsection{Intention Understanding}
In task-oriented semantic communications, the user wants the receiver to perform the desired task. Recognizing the users' intentions is the key part of performing the task successfully, especially for the multi-task scenario in which multiple tasks are performed simultaneously. The new semantic communication architecture can help the receiver recognize the users' intentions. The context information from the transmitter generally contains the users' intentions, where the intentions could be performing a simple task as well as a series of tasks. We can use the memory module to store and process the received context information first, then provide it to the large language models (ChatGPT, LLamDA, etc.) to understand the user's intention clearly, and finally choose the appropriate model to perform the task. As shown in Fig.~\ref{fig:applications}(a), the user transmits the sentence to the receiver first, then the receiver recognizes the intention that is to perform the image classification task, and finally it chooses the corresponding model to classify the object in the received image. 

\subsection{Multimodal Generation}
This application uses large models, e.g., text-to-image or text-to-video models (DALL·E, Midjourney, Stable Diffusion, etc.), for generating multimodal data. The personalized services, e.g., metaverse and personalized lives, will be beneficial from this application with lower latency and more accurate content generation. As shown in Fig.~\ref{fig:applications}(b), the large models can use the received semantics to generate artist-level videos or images, which extends the capability of semantic communication from data reconstruction to personalized data generation. The memory module can provide additional short-term to control the data generation. For example, given the sentence ``\textit{Generate a rabbit playing guitar video, modern Disney style.}''  the memory module can provide the rabbit image received from the user in the last time slot, and the large model can offer information about playing guitar and Disney style. Remark that rabbit images belong to short-term knowledge since they could vary from the users, however playing guitar and Disney style are the common knowledge for most users. Compared with conventional communications to transmit the sentence directly, large model-empowered semantic communications can convey the semantics more accurately thus achieving the desired data generation.  

\subsection{Complex Tasks}
The proposed semantic communication architecture enables semantic communications to perform more complex tasks. The large models have the capacity to decompose a complex task into multiple simple problems through the chain of thought and invoke different semantic decoders to finish the complex task jointly. Moreover, since the large model belongs to the multitask model, it is possible to replace the multiple semantic decoders with only one large model. With the memory module, the systems can perform both memoryless tasks and memory tasks. Memoryless tasks are only relevant to the inputs received in the current time slot, e.g., receiving the image and recognizing its category. Memory tasks are relevant to inputs received in both the current and past time slots, e.g., the response in the conversation relying not only on the currently listened sentences but also on the previous context. As shown in Fig.~\ref{fig:applications}(c), consider the scenario visual question answer task, the memory module can store the semantics of the received images and then provide them to the large models. Given the question, ``\textit{How many balls in this picture and last picture?}'' The large model can understand the intention first and then decompose it into several steps, i.e., identifying the balls in the image and counting the number.

\subsection{Scenario Generation}
The application demonstrates that large models can improve the robustness of semantic communications. Collecting real-world data is challenging, but the generatively large model can generate the training data for both the communication and semantic aspects. {For example, in autonomous driving, generative AI offers more diverse and automated data, e.g., semantic multi-view panoramas and reconstructed real scenes, providing a wealth of labeled data for model training.}

In cases where collecting data for new communication scenarios is arduous, {the generatively large model can simulate the communication environment at the edge with the simulator descriptions.} Then, the transceiver can be trained with the generated scenario first such that to adopt the new environment, which helps semantic communication systems adapt to different communication environments quickly. The example is shown in Fig.~\ref{fig:applications}(d), with the scenario descriptions, the generated channels can be employed for training the image semantic communication in an end-to-end manner. Furthermore, the generative scenarios offer valuable supplementary information for optimizing beamforming, which therefore improves the robustness of transmission. Beyond their utility in communications, these generative large models extend their versatility to generate diverse data types for various tasks. For example, it can generate a large number of images with specific categories for the image classification task and image retrieval.

{These applications can be applied to real-world scenarios by deploying the large models locally, e.g., pedestrian re-identification, scenario tasks, robot control, and by cooperating with the large models at edge/cloud, e.g., simulation environment generation, extended reality (XR), and virtual reality (VR). For example, the new architecture can transmit accurate semantics to edge and then feedback the generated virtual items in VR.}

\begin{figure*}[!t]
    \centering
    \includegraphics[width=180mm]{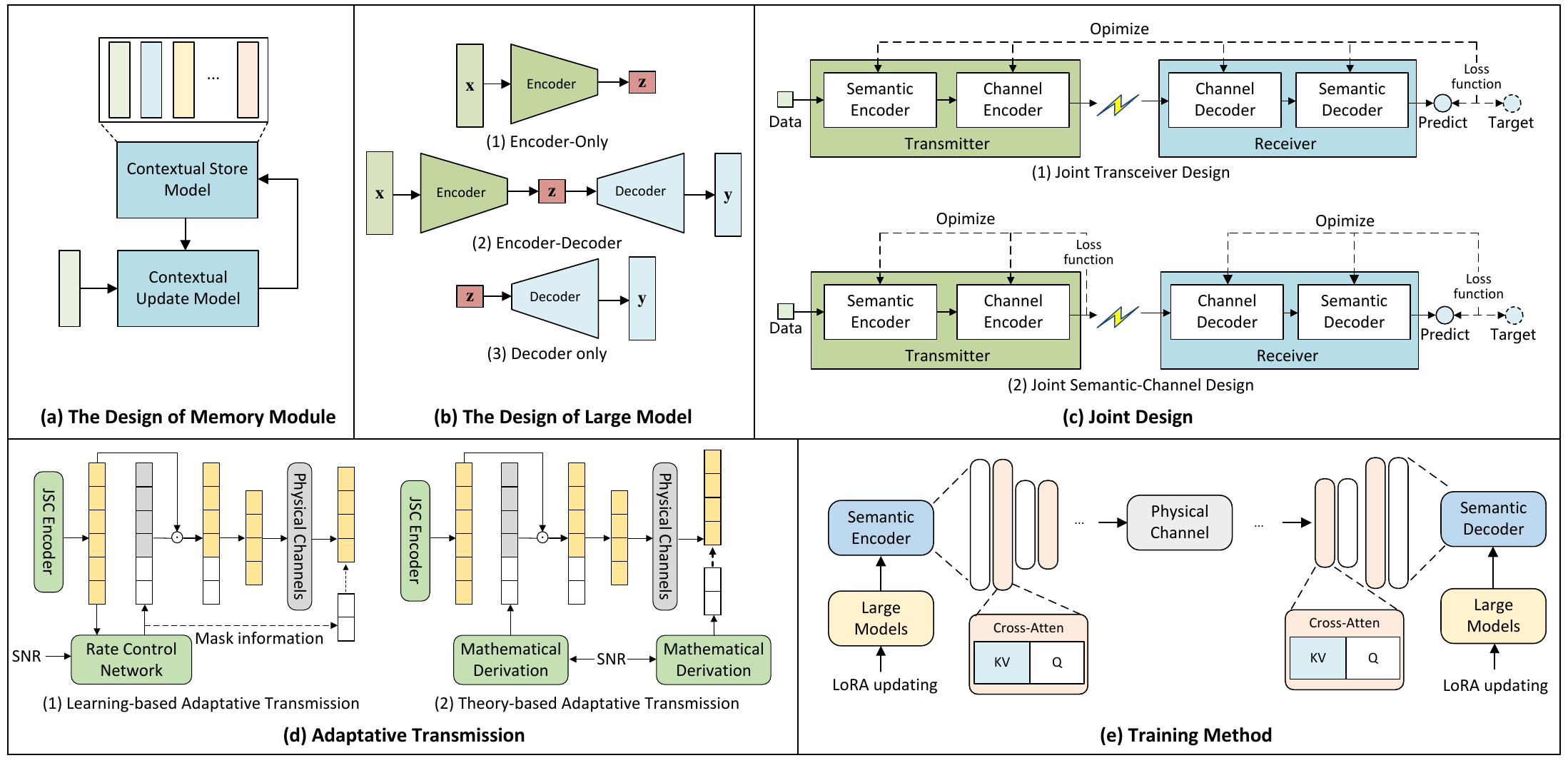}
    \caption{The key designs of memory module, large model, joint training, and adaptive transmission.}
    \label{fig:key-designs}
\end{figure*}

\begin{figure}[!t]
    \centering
    \includegraphics[width=70mm]{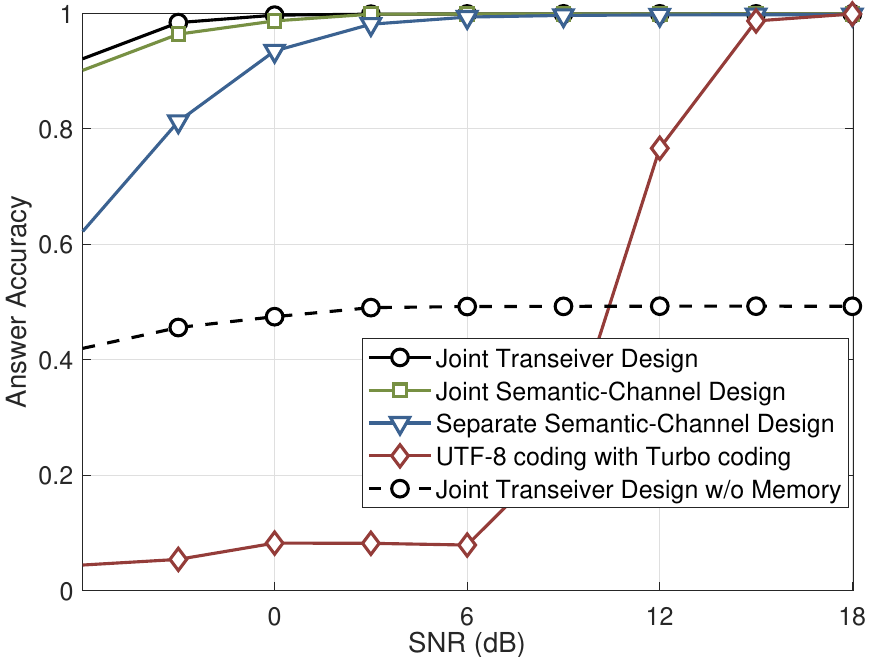}
    \caption{Answer accuracy comparison between semantic communication systems over AWGN channels. }
    \label{fig:memory-module}
\end{figure} 

\section{Key Designs}\label{sec-iv}
Besides the typical applications, in this section, we present the main challenges in designing and implementing large model-empowered semantic communications, which have the design of memory modules, the design of large models, the design of joint training, the adaptive transmission, and the training method.

\subsection{The Design of Memory Module}
The design of the memory module mainly focuses on the contextual store model and contextual update model, which is shown in Fig. \ref{fig:key-designs}(a).
\subsubsection{Contextual Store Model}
The design of the contextual store model needs to consider the size of the space, the alignment of multimodally contextual information, and the design of temporal coding. The size of space affects how much contextual information can be stored, where the smaller space fits the tasks that require fresh information and low computation power, and vice versa. In addition, the alignment of multimodally contextual information is to find relationships and connections between two or more modalities. A better alignment design can help improve the understanding of multimodal contextual information and make context management easier. Moreover, for contextual information, it is important to recognize the order of features in the context that happened earlier or later. Thus, we need to design the appropriate temporal coding to describe the temporal relationship between these contextual features. 

\subsubsection{Contextual Update Model}
The contextual update model should carefully design the updating rules to replace the context with the incoming context. The rules could be the freshness of information, the value of information, the correlation of information, and so on. The freshness of information is for the tasks demanding frequent and regular updates of certain information. Timely updates of the context are an important aspect of such tasks, e.g., metaverse,  remote control/monitoring of autonomous vehicles, etc. The value of information measures the contributions of the context to the tasks, which is suitable for resource allocation decisions.  The measurement of correlations between contexts is beneficial for tasks requiring long-term retrieval, in which the least relevant context will be replaced. These rules could be used separately or jointly to update the stored context, such that provide more accurate contextual information for the tasks.

The recent work \cite{10159023} is to formulate the memory module as a queue with finite length for memory tasks. In Fig.~\ref{fig:memory-module}, we show the comparison for the semantic communications with and without the memory module. The scenario question-answer task is considered. It is interesting to observe that the semantic communication system with a memory module outperforms that without a memory module significantly in terms of answer accuracy.

\subsection{The Design of Large Model}
The design of large models can be divided into encoder-only, decoder-only, and encoder-decoder,  which are shown in Fig. \ref{fig:key-designs}(b). These designs focus on different capabilities, i.e., understanding and generation. 

\subsubsection{Encoder-only Design}
The encoder-only model mainly focuses exclusively on encoding input data into a fixed-dimensional representation, often referred to as embeddings or latent representations, which is to capture meaningful features or representations of the input data. This type of model is trained by masking the information randomly and predicting the masked information with the unmasked information. The absence of a decoder in encoder-only models simplifies the architecture and reduces computational complexity, making them efficient for tasks that require feature extraction or representation learning,  e.g., classification tasks, without the need for data generation or sequence-to-sequence transformations. The representative works of the encoder-only model are BERT for text and Mask-Autoencoder for the image. 

\subsubsection{Decoder-only Design}
The decoder-only model, also called the generative model, focuses exclusively on generating output data or sequences from given inputs or latent representations, which is to transform input data or latent representations into target sequences or structured outputs. The type of model is trained in an auto-aggressive manner generally, predicting the next word or pixels based on the previous words or pixels. The absence of an encoder in decoder-only models weakens the understanding ability but enhances the generative ability, which is suitable for tasks that require data generation or sequence-to-sequence transformations. GPT-4 and stable diffusion models are the representative works.

\subsubsection{Encoder-Decoder Design}
The encoder-decoder model employs the encoder to learn the latent represents of source data and the decoder to generate the target data, in which the encoder represents the understanding ability and the decoder has the generative ability. This type of model can be trained using supervised learning, where they learn to map input sequences to target sequences using pairs of aligned data (e.g., source and target sentences in machine translation). This design considers data understanding and data generation together and can be adapted for tasks like image captioning, where the encoder processes an image, and the decoder generates a textual description. The recent works of the encoder-decoder model are Flan-UL2 and Flan-T5. 

\begin{figure}[!t]
    \centering
    \includegraphics[width=70mm]{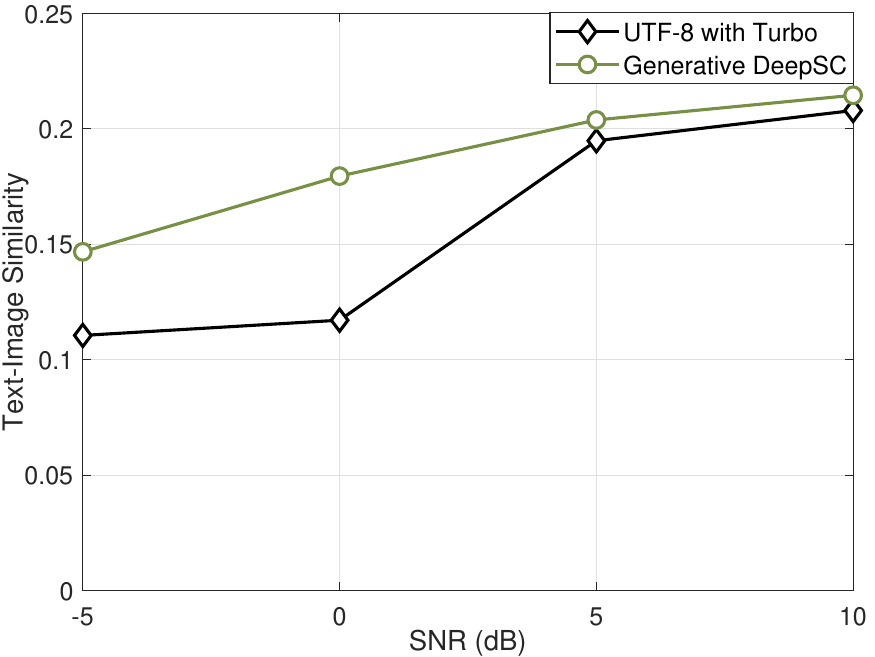}
    \caption{Text-image similarity comparison between generative DeepSC and conventional method over AWGN channels }
    \label{fig:generative-results-1}
\end{figure}

\begin{figure}[!t]
    \centering
    \includegraphics[width=90mm]{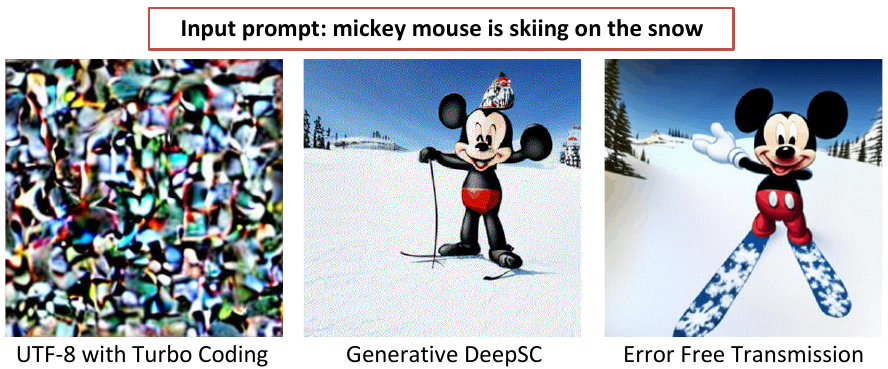}
    \caption{Visualized results of text-to-image task over AWGN channels. }
    \label{fig:generative-results-2}
\end{figure}

The recent works \cite{tay2022ul2, RombachBLEO22, thoppilan2022lamda} mainly adopt the decoder-only design due to the amazingly generative results and more efficient running on capacity-limited devices.  In Fig.~\ref{fig:generative-results-1}, we show the text-to-image similarity comparison between generative DeepSC and UTF-8 with turbo coding over AWGN channels. The generative DeepSC consists of the DeepSC-MT and the large diffusion model to generate images based on text semantic information, in which the DeepSC-MT and large diffusion model are trained jointly.  The generative DeepSC outperforms the conventional methods in low SNR regimes. From Fig.~\ref{fig:generative-results-2}, we can observe that the generative DeepSC can generate more accurate images than the conventional method. 

\vspace{-1em}

\subsection{Joint or Separate Design}
Based on Shannon's theorem, current communication systems adopt a separate design, where each module is optimized separately. Such a separate design does not consider the error propagation between each module, which achieves local optimization. Therefore, we can jointly design and train the model to alleviate the error propagation. There exist two kinds of joint design, shown in Fig. \ref{fig:key-designs}(c), 1) the joint transceiver design and 2) the joint semantic-channel coding design. 

\subsubsection{joint transceiver training}
First, we discuss the joint transceiver training. By feeding the large training data and setting loss functions, the weights of the transmitter and receiver will be updated by the stochastic gradient descent (SGD) algorithm with the back-propagation from receiver to transmitter, which can alleviate the error propagation and achieve global optimization in an end-to-end manner. However, it needs a stable feedback channel to transmit the gradients accurately for online training. Therefore, such design generally is taken locally and then deployed the models to devices. 

\subsubsection{joint semantic-channel coding design}
The second one is the joint semantic-channel coding design, in which the semantic coding and channel coding are optimized jointly but the transmitter and receiver are trained separately. This design does not need gradient feedback from the receiver to the transmitter, which is suitable for online training. Due to the local optimization, the joint semantic-channel coding design slightly underperforms the joint transceiver design in some cases but outperforms the separate design. 

Choosing which joint design depends on the availability of stable feedback channels.  The recent works of semantic communications \cite{9837870, DaiWTSQ0022, 10159023}  mainly adopt the joint transceiver design due to the better performance in terms of data reconstruction and task execution. In Fig.~\ref{fig:memory-module}, we show the comparison for different designs. The simulation results demonstrate that the joint design can achieve a better answer accuracy than the separate design.

\subsection{Adaptive Transmission}
The adaptive transmission in conventional communication can adopt different combinations of channel coding rates and modulation orders to avoid outages for different channel conditions. In semantic communications, adaptive transmission can be achieved by masking the unessential elements in the transmitted signals, i.e., masking less at low signal-to-noise ratio (SNR) regimes to ensure the reliability of performing tasks and masking more elements at high SNR regimes to achieve a higher transmission rate. As shown in Fig.~\ref{fig:key-designs}(d), the design of adaptive transmission in semantic communication has two types: 1) learning-based and 2) theory-based. 

\subsubsection{The Learning-based Adaptive Transmission} 
The learning-based method introduces the rate control network to predict the number of masked elements for different SNRs by adversarial training, where the number of mask elements and the quality of data reconstruction achieve the Nash equilibrium. However, this learning-based method faces several challenges. First, this method may face mode collapse, in which the number of masked elements for different SNRs will converge to the same. A well-designed network structure and hyper-parameters are required. Besides, since the mask generation is based on both the dynamic semantic information and the SNRs, the mask information should be transmitted to the receiver to pad zeros. Different rate control networks have been proposed, i.e., the spatial mask network and the Gumbel softmax-based \cite{yang2021deep}, and so on.

\subsubsection{The Theory-based Adaptive Transmission}
The relationship between the number of masked elements and SNRs is derived mathematically~\cite{10159023}. Then, the transceiver is trained by masking the pre-defined number of elements. The theory-based method can avoid the mode collapse and does not need to transmit the mask information. However, it needs to find the connections between semantic noise and channel noise. Sometimes, semantic noise is hard to model mathematically, especially for large models.

\subsection{Training Method}
The large model includes billions of parameters, making training processing time-consuming and computation-intensive. Jointly training the system with the large models becomes less feasible. As shown in Fig.~\ref{fig:key-designs}(e), instead of updating full parameters, we can employ the Low-Rank Adaptation (LoRA)~\cite{HuSWALWWC22} to update the parameters of a large model, where the gradients are replaced with two learnable low-rank matrices of weights. {We can also use LoRA with federated training to protect user's privacy.} Besides, for training semantic codec, we can freeze the large model and introduce the cross-attention layer to merge the knowledge from the large model.

\section{Conclusion and Future research directions}\label{sec-v}
This article proposes large model-empowered semantic communications to fully utilize the power of a large model to enhance the capacity of semantic communication. The new modules have been developed to support the large model-empowered semantic communication systems, the following challenges should be addressed:

\subsection{The Modeling of Memory Module}
The current memory module is only modeled as the queue with finite length following the first-in first-out, in which the memory store model is the queue and the memory update model is the first-in first-out. However, such a design does not consider the alignment of multimodal semantics and the correlation between context information, therefore it is not suitable for more complex scenarios. To support more effective processing of contextual information, we need to design novel schemes on the memory store model and memory update model to maximize the utilization efficiency.

\subsection{The Deployment of Large Model}
Even though the large model has shown powerful capacities for various tasks, it includes hundreds of billions of parameters, which makes the large model hard to deploy on mobile devices. {The cost of applying large models to semantic communications directly is the high power consumption and even device crashes.} One way to deploy the large models to the capacity-limited device is to compress the large model by model compression techniques, i.e., weights pruning, weights quantization, and knowledge distillation. {Another way is to perform large models with edge computing, which can offload the computation-intensive tasks to the edge server. However, compression techniques and edge computing will inevitably degrade the performance of large models or increase latency.} Therefore, new model compression and edge computing techniques to achieve graceful performance degradation are urgently needed to speed up the deployment of the large model by considering the trade-off between computation resource, communication resource, and performance.

\subsection{The Joint Training Algorithm}
Training the large model from scratch is money-consuming and time-consuming. For example, GPT-3 is trained more than 1,000,000 V100 GPU hours. Thus, the mainstream is to train the base model with billions of parameters and then fine-tune the pre-trained base model for the downstream tasks with some epochs. However, such fine-tuning is still time-consuming and is not stable, which depends on the training data and the settings of hyper-parameters. Besides, semantic communication systems are generally trained in a joint manner for better performance. The additional modules, i.e., memory module and channel codec, will also increase the complexity of joint training. Therefore, it is still essential to design an effective training algorithm to speed up joint training of the pre-trained large model, memory module, and channel codec.

\subsection{The Multimodal Information Processing}
The recent works focus on the two modality information processing, i.e., text and image, which has shown superiority over the model trained with only one modality. The real world exists not only text and image, but also the other modalities of data, e.g., audio, depth-image, radar information, CSI information, etc. Besides, the memory module introduces multimodal information in the time domain. Merge all the new types of data can improve the generalization of large models for more tasks in the real world. How to design the model to fuse this information remains to be studied.

\appendices

\ifCLASSOPTIONcaptionsoff
  \newpage
\fi

\bibliographystyle{IEEEtran}
\bibliography{bibtex/reference}

\vspace{-1cm}
\begin{IEEEbiographynophoto}{Huiqiang Xie} is an Associate Professor at Jinan University.  He received 2023 IEEE ICC Best Paper Award and 2023 IEEE SPS Best Paper Award.
\end{IEEEbiographynophoto}
\vspace{-1cm}

\begin{IEEEbiographynophoto}{Zhijin Qin} is an Associate Professor at Tsinghua University. She served as a guest editor and IEEE JSAC special issue on semantic communications and an associate editor of IEEE Trans. Communications. She has received several awards, including 2017 IEEE GLOBECOM Best Paper Award, 2018 IEEE Signal Processing Society Young Author Best Paper Award, 2021 IEEE Communications Society SPCC Early Achievement Award,  2022 IEEE Communications Society Fred W. Ellersick Prize,  2023 IEEE ICC Best Paper Award, and  2023 IEEE SPS Best Paper Award.
\end{IEEEbiographynophoto}

\vspace{-1cm}
\begin{IEEEbiographynophoto}{Xiaoming Tao} is a Professor at Tsinghua University. She served as a workshop general co-chair for IEEE INFOCOM 2015, and the volunteer leadership for IEEE ICIP 2017. She is serving as the Associate Editor of IEEE Trans. Wireless Communications and several other journals. She is also the recipient of National Science Foundation for Outstanding Youth and many national awards, such as 2017 China Young Women Scientists Award, 2017 First Prize of Wu Wen Jun AI Science and Technology Award, and 2016 National Award for Technological Invention Progress
\end{IEEEbiographynophoto}

\vspace{-1cm}
\begin{IEEEbiographynophoto}{Zhu Han}  is a professor in the Electrical and Computer Engineering Department and the Computer Science Department at the University of Houston, Houston, Texas 77004 USA. He received an NSF Career Award in 2010, the Fred W. Ellersick Prize of the IEEE Communication Society in 2011, the EURASIP Best Paper Award for the Journal on Advances in Signal Processing in 2015, IEEE Leonard G. Abraham Prize in the field of Communications Systems (Best Paper Award in IEEE JSAC) in 2016, IEEE Vehicular Technology Society 2022 Best Land Transportation Paper Award, and several best paper awards in IEEE conferences.
\end{IEEEbiographynophoto}

\end{document}